\documentclass[
	a4paper, 
	8pt, 
]{LTJournalArticle}

\DeclareUnicodeCharacter{0301}{\'{e}}
\usepackage{etoolbox}
\providetoggle{preprint}
\settoggle{preprint}{true}

\usepackage{amsmath,amssymb,amsfonts}
\usepackage{mathtools}
\usepackage{algorithmic}
\usepackage{graphicx} 
\usepackage[T1]{fontenc}
\usepackage{booktabs}
\usepackage{authblk}
\usepackage{multirow}
\usepackage{tabularx}
\usepackage{pgf}
\usepackage{tikz} 
\usetikzlibrary{spy,math,arrows}
\usepackage{hyperref}
\hypersetup{
	colorlinks = true,
	linkcolor=black,
	citecolor={blue!50!black},
	urlcolor={blue!80!black},
	anchorcolor = blue
}
\hypersetup{linkcolor=black}

\usepackage{xspace}
\usepackage{bm}
\usepackage{cleveref}
\newcommand{\vx}{\mathbf{x}}
\newcommand{\vy}{\mathbf{y}}
\newcommand{\vz}{\mathbf{z}}
\newcommand{\vc}{\mathbf{c}}
\newcommand{\ve}{\mathbf{e}}
\newcommand{\vf}{\mathbf{f}}
\newcommand{\vm}{\mathbf{m}}
\newcommand{\vs}{\mathbf{s}}
\newcommand{\vp}{\mathbf{p}}
\newcommand{\vb}{\mathbf{b}}
\newcommand{\PC}{\mathrm{PC}}
\newcommand{\LD}{\mathrm{LD}}
\newcommand{\SD}{\mathrm{SD}}
\newcommand{\CE}{\mathrm{CE}}
\newcommand{\R}{\mathbb{R}}

\newcommand{\mW}{\mathbf{W}}

\newcommand{\sigmoid}{\mathrm{sigmoid}}

\definecolor{highlight}{HTML}{FEE34E}
\tikzset{
image/.style={inner sep=0},
marker/.style={circle, draw=highlight, thick, inner sep=0pt,minimum size=3mm, draw opacity=.75},
arrow/.style={draw=highlight, >=latex, thick, draw opacity=.75},
}

\newcommand\copyrighttext{%
  \footnotesize This work has been submitted to the IEEE for possible publication. Copyright may be transferred without notice, after which this version may no longer be accessible.}
\newcommand\copyrightnotice{%
\begin{tikzpicture}[remember picture,overlay]
\node[anchor=south,yshift=1pt] at (current page.south) {\fbox{\parbox{\dimexpr\textwidth-\fboxsep-\fboxrule\relax}{\copyrighttext}}};
\end{tikzpicture}%
}

\begin{document}

\title{Gadolinium dose reduction for brain MRI using conditional deep learning}

\author[1,*]{Thomas Pinetz}
\author[2,*]{Erich Kobler}
\author[2]{Robert Haase}
\author[3]{Julian A. Luetkens}
\author[4]{Mathias Meetschen}
\author[4]{Johannes Haubold}
\author[4]{Cornelius Deuschl}
\author[2]{Alexander Radbruch}
\author[2]{Katerina Deike}
\author[1]{Alexander Effland}

\affil[1]{Institute for Applied Mathematics, University of Bonn}
\affil[2]{Department of Neuroradiology, University Hospital Bonn}
\affil[3]{Diagnostic and Interventional Radiology, University Hospital Bonn}
\affil[4]{Diagnostic and Interventional Radiology, University Hospital Essen}

\date{\vspace{-5ex}}

%
%
%
%
\maketitle              

\begin{abstract}

Recently, deep learning (DL)-based methods have been proposed for the computational reduction of gadolinium-based contrast agents (GBCAs) to mitigate adverse side effects while preserving diagnostic value.
Currently, the two main challenges for these approaches are the accurate prediction of contrast enhancement and the synthesis of realistic images.
In this work, we address both challenges by utilizing the contrast signal encoded in the subtraction images of pre-contrast and post-contrast image pairs.
To avoid the synthesis of any noise or artifacts and solely focus on contrast signal extraction and enhancement from low-dose subtraction images, we train our DL model using noise-free standard-dose subtraction images as targets.
As a result, our model predicts the contrast enhancement signal only; thereby enabling synthesization of images beyond the standard dose.  
Furthermore, we adapt the embedding idea of recent diffusion-based models to condition our model on physical parameters affecting the contrast enhancement behavior.
We demonstrate the effectiveness of our approach on synthetic and real datasets using various scanners, field strengths, and contrast agents.
\end{abstract}

\copyrightnotice

\section{Introduction}
\label{sec:introduction}

\begin{figure}[t]
    \centering
    \includegraphics[width=\linewidth]{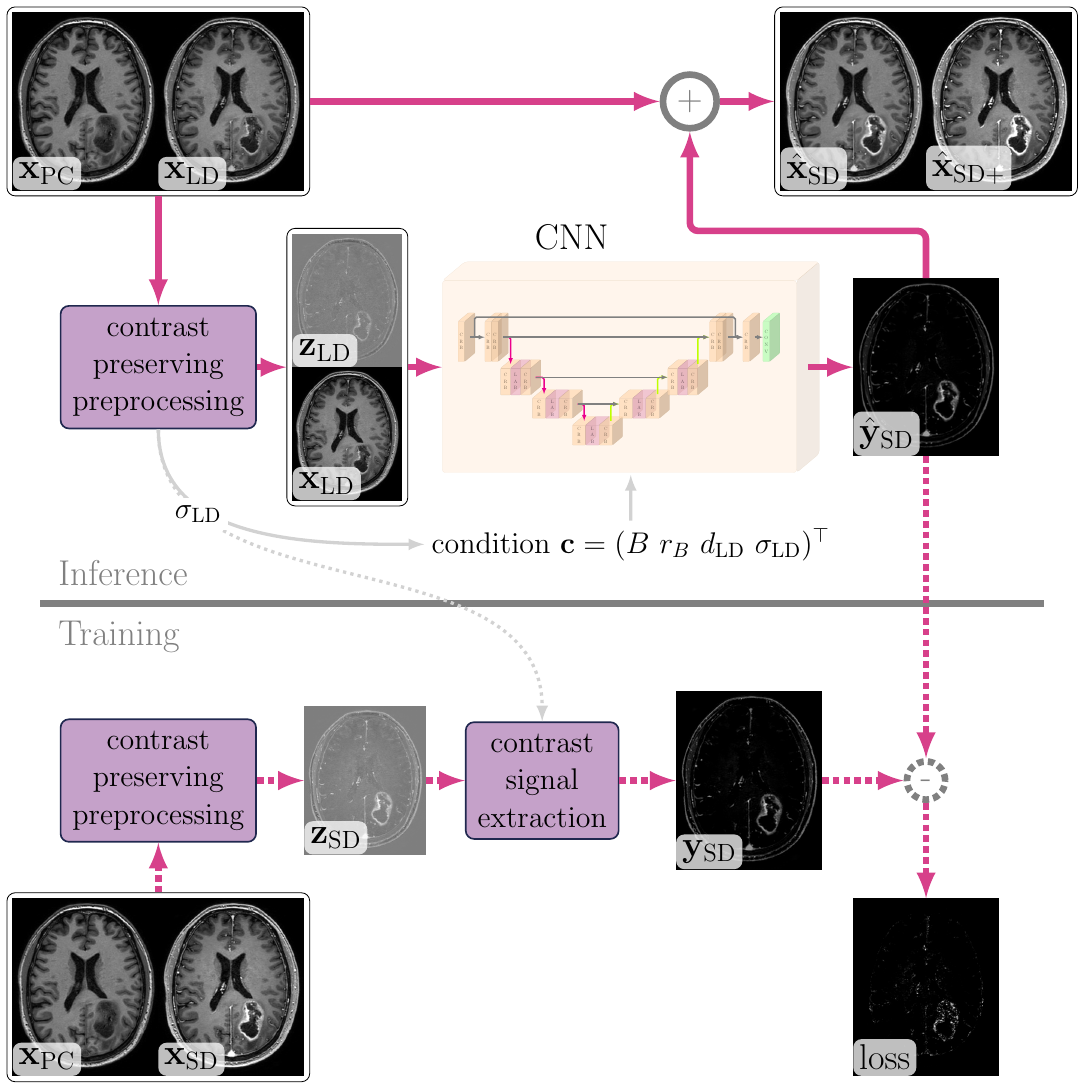}
    \caption{Visualization of our dose reduction and contrast enhancement DL approach.
    The inference process is illustrated at the top using solid arrows, while the additional processing steps for training are shown at the bottom.
    In both cases, pairs of pre-contrast~\(\vx_\PC\) and contrast-enhanced images~\(\vx_{\{\LD,\SD\}}\) are preprocessed to obtain subtraction images~\(\vz_{\{\LD,\SD\}}\).
    The pairing of low-dose subtraction and low-dose image~$\{\tilde{\vz}_\LD,\vx_\LD\}$ is then processed by a conditional CNN generating a contrast signal~\(\hat{\vy}_\SD\) that approximates the target~\(\vy_\SD\). This target is obtained by removing the noise from the standard-dose subtraction image~\(\vz_\SD\).
    At inference the predicted contrast signal~\(\hat{\vy}_\SD\) is added to the pre-contrast~\(\vx_\PC\) or low-dose image~\(\vx_\LD\) to obtain a synthesized standard-dose image~\(\hat{\vx}_\SD\) or a further contrast enhanced image~\(\hat{\vx}_{\SD+}\).}
    \label{fig:overview}
\end{figure}

Gadolinium-based contrast agents (GBCAs) are indispensable in first-line diagnosis and follow-up imaging of various neuropathologies ranging from multiple sclerosis to metastasis and tumors.
However, GBCAs are costly, may cause patient discomfort, can accumulate in the body independent of kidney function~\cite{KaIs14}, and have a potential risk of causing nephrogenic systemic fibrosis in patients with impaired renal function~\cite{ScBl18}.
Furthermore, the gadolinium administered during magnetic resonance imaging (MRI) examinations accumulates in the drinking water system~\cite{BuHo20}.
Therefore, current guidelines recommend administering the lowest dose of GBCA required for a reliable diagnostic result~\cite{ACR22}.

In recent years, deep learning has had a prominent impact on diagnostics in medicine and radiology in particular.
Currently, major MRI vendors offer deep-learning (DL) solutions to common imaging problems, such as denoising, super-resolution, and motion correction~\cite{LiWa23}.
However, DL image generation approaches tend to create unrealistically smooth images.
In the worst case, this could result in pathologies becoming invisible.
To find a tradeoff between noise removal and small structure/texture preservation, distributional losses such as perceptual or adversarial losses~\cite{JoAl16, GoPo14, PaJo21} or local Wasserstein distances~\cite{PiKo23} have been used.
These losses are known to cause hallucinations~\cite{CoMa18, AnRe20}, as the generative models need to synthesize noise and artifacts to match the statistics of real MRI images.
In turn, these artifacts or noise pixels can be interpreted as small pathologies~\cite{HaPi23}, which could result in misdiagnosis.

Recent DL-based contrast agent dose reduction methods in MRI learned to predict a synthetic contrast-enhanced (CE) standard-dose image~\(\hat{\vx}_\SD\) from at least a pre-contrast~\(\vx_\PC\) and low-dose CE~\(\vx_\LD\) image pair~\cite{GoPa18,PaJo21,AmBo22}.
Their deep convolutional neural networks (CNNs) were trained to directly approximate the corresponding standard-dose image~\(\vx_\SD\) and are therefore prone to the previously discussed tradeoff.
In contrast, we propose to disentangle contrast signal enhancement from the synthesis of realistic standard-dose images.
To this end, we first extract the noise and contrast signal from a low-dose image by subtracting the corresponding pre-contrast image, i.e., acquired before administering any GBCA, as illustrated in Figure~\ref{fig:overview}.
The CNN processes the resulting low-dose subtraction image to remove the noise and enhance the contrast signal.
An image equivalent to the standard-dose image is obtained by adding the predicted contrast signal~\(\hat{\vy}_\SD\) to the pre-contrast image~\(\vx_\PC\).
We train the CNN by penalizing the deviation of its output from a contrast enhancement image~\(\vy_\SD\).
This target image is computed by extracting the CE regions from the standard-dose subtraction image using a simple statistical model suppressing noise.
The focus on subtraction images enables the synthesis of CE images beyond the standard dose by adding the predicted CE image~\(\hat{\vy}_\SD\) to the low-dose image~\(\vx_\LD\) instead of the pre-contrast image~\(\vx_\PC\).


Various physical factors such as the scanner's field strength as well as the dose and relaxivity of the administered contrast agent influence the general enhancement behavior within CE images~\cite{ShGo15, KuMa22, MiSe23}.
However, this has not been explored by recent DL-based dose reduction methods due to the lack of multi-center data for the evaluation of domain shifts.
We propose to explicitly condition the CNN onto these physical quantities for improved generalization.
In particular, we transferred the conditioning approach of denoising diffusion models~\cite{HoJa20} to our setting.
Using a non-linear embedding of these physical factors, each layer of our CNN can be adapted, thus promoting visual analysis of the learned embedding and increasing the model's interpretability.

To assess the efficacy of our approach, we perform a qualitative and quantitative comparison to recent DL-based dose reduction methods.
More precisely, we used the openly accessible brain tumor segmentation metastasis (BraTS-METS) dataset~\cite{MoJa23} to synthesize low-dose CE images using~\cite{PiKo23}.
In addition, we train and test all methods on an in-house dataset containing real low-dose CE images at different dose levels.
For testing the generalization to different settings, we further compiled a dataset from an external site exhibiting diverse pathologies.
The numerical results on all datasets show that our approach outperforms recent approaches qualitatively and quantitatively.

\section{Related Work}

To account for the as low as reasonable possible dosage recommendations, various gadolinium dose reduction techniques have been proposed.
While initial research has focused on developing new contrast agents with increased relaxivity~\cite{RoPo19}, the advances of deep learning~\cite{LeBe15} enabled a complementary research direction called virtual contrast~\cite{HaPi23b, MoEl23}.
In these approaches, an contrast-enhanced T1-weighted (T1w) image with substantially reduced GBCA dose is acquired along with pre-contrast images such as T1w, T2w, or FLAIR.
These images are preprocessed and fed into a deep neural network to extract the subtle CE signal and synthesize an artificial CE T1w image resembling a corresponding standard-dose image.

In 2018, Gong et al.~\cite{GoPa18} introduced the first virtual dose reduction approach that combined a 2D U-Net~\cite{RoFi15} with residual learning to reduce the GBCA dosage by~$90\%$.
Their model was limited to processing each 2D slice individually.
Thus, no information across slices was processed. 
Later, Pasumarthi et al.~\cite{PaJo21} introduced a 2.5D approach by feeding 7 neighboring slices into the U-Net-like model to extract relevant information from adjacent slices.
To further exploit 3D information, they performed a multi-planar reconstruction approach, in which predictions for different orientations and rotations of the input images were computed and the pixel-wise average defined the model output.
In addition, they used a combination of perceptual and adversarial losses to counteract a smooth appearance of the synthesized images, which is advocated by the $\ell_1$ and structured similarity index loss functions originally used in~\cite{GoPa18}.

In parallel, B\^{o}ne et al.~\cite{BoAm21} developed a fully 3D U-Net-like CNN for a dose reduction factor of $75\%$.
In contrast to the previous two approaches, B\^{o}ne et al. processed the images using 3D convolution layers and residual blocks in their architecture.
Their model was trained to synthesize standard-dose images that minimize the $\ell_2$-distance to the corresponding standard-dose images from 4 input sequences (pre-contrast T1w, low-dose T1w, FLAIR, ADC).

To evaluate the medical benefit, several reader studies were performed:
despite the impressive image quality of the synthesized CE images, significant performance issues in small and weakly-enhancing lesions~\cite{LuZh21, AmBo22, HaPi23} as well as hallucinations of lesions~\cite{AmBo22, HaPi23} limited clinical applicability. 
In addition, the resulting synthesized images are severely smoothed and could be distinguished from the real ones~\cite{AmBo22}.

All previously discussed methods were trained on datasets that consist of matched pre-contrast, low- and standard-dose scans.
The compilation of such datasets is time- and resource-intensive since low-dose CE images are not acquired in clinical practice.
To avoid the acquisition of real low-dose images, recently methods synthesizing low-dose images from pre-contrast and standard-dose image pairs were proposed.
Mingo et al.~\cite{MiSe23} proposed an approximative physical model to generate half-dose images by averaging pre-contrast and standard-dose images. 
In contrast, Pinetz et al.~\cite{PiKo23} facilitated generative adversarial networks to synthesize low-dose images at various dose levels.
To faithfully predict the noise and low-dose contrast signal, they proposed a novel perceptual loss function based on the Wasserstein distance of local patches.
They evaluated the benefit of synthetic low-dose images for training the model of Pasumarthi et al.~\cite{PaJo21}.
It turned out that training on just synthetic low-dose images is inferior compared to real-world low-dose images.
However, augmenting the training data by synthetic low-dose images from different sites led to improved results.
Similarly, Wang et al.~\cite{WaPa23} proposed to synthesize low-dose CE images from pre-contrast and standard-dose image pairs using an iterative transformer model.
They considered a global attention mechanism on subsampled feature maps in combination with rotational shifts to focus the model on the overall contrast uptake behavior.
They evaluated the efficacy of the synthetic low-dose images for the downstream tasks of standard-dose image synthesis and subsequent tumor segmentation.

A different line of research~\cite{KlMo19, PrMe21, LiPa23} focuses on synthesizing CE standard-dose images from just pre-contrast images, i.e., without administering any GBCA contrast agent.
The study of Preetha et al.~\cite{PrMe21} demonstrated that synthetic CE standard-dose images could potentially be used for tracking tumor progression.
However, the performance of these methods is not sufficient for more general clinical settings~\cite{HaPi23b}.
Furthermore, Liu et al.~\cite{LiPa23} found that synthesizing CE T1w images from pre-contrast images only is a hard problem and significantly deteriorates automatic tumor segmentation.


\section{Method}

In this work, our main goal is to extract and enhance the differential contrast signal encoded in subtraction images of a contrast-enhanced (CE) T1w image and the corresponding \emph{pre-contrast} T1w image~\(\vx_\PC\in\R^n\) (i.e. acquired before injecting any contrast agent), where \(n\) denotes the number of pixels or voxels.
More precisely, we consider CE images that are acquired after administering the \emph{standard-dose}~\(\vx_\SD\in\R^n\) or a \emph{lower-dose}~\(\vx_\LD\in\R^n\).
We denote the administered dose of a CE scan by~\(d\) and set the weight-dependent standard dose to~\(d_\SD=1\).
Thus, low-dose CE images~\(\vx_\LD\) are acquired with dose levels~\(d_\LD\in(0,1)\).

Our approach consists of three essential concepts, which are related in Figure~\ref{fig:overview}.
First, a contrast-preserving preprocessing yields subtraction images~\(\vz_\LD=\vx_\LD-\vx_\PC\) and \(\vz_\SD=\vx_\SD-\vx_\PC\).
Second, a conditional convolution neural network (CNN) removes noise and artifacts from the low-dose subtraction image~\(\vz_\LD\) and predicts the contrast signal~\(\hat{\vy}_\SD\).
This network is conditioned on metadata to account for physical parameters that affect the general contrast enhancement behavior.
Besides the relative administered dose~\(d_\LD\), we consider the field strength of the scanner~\(B\in\{1.5,3\}\) and the field strength-dependent relaxivity of the used contrast agent~\(r_B\in\R_+\).
In addition, we include an estimate of the noise level~\(\sigma_\LD\) of the low-dose subtraction image~\(\vz_\LD\) motivated by the recent success of diffusion models~\cite{HoJa20}.
Third, during training we process \(\vz_\SD\) to extract the contrast signal image~\(\vy_\SD\), which only contains the noise-free contrast signal.

Once the CNN is trained it predicts a noise-free contrast enhancement signal image~\(\hat{\vy}_\SD\in\R^n\).
A particular advantage of this approach is that this enhancement signal image can be added to the pre-contrast \emph{or} low-dose image to generate anatomical images~\(\hat{\vx}_\SD\in\R^n\) or \(\hat{\vx}_{\SD+}\), as shown in the top right of Figure~\ref{fig:overview}.
Thus, we can use our model simultaneously for GBCA dose reduction and contrast signal enhancement beyond standard-dose images.

\subsection{Contrast-preserving preprocessing}
\label{sec:Preprocessing}

\begin{figure}[tb]
    \includegraphics[width=\linewidth]{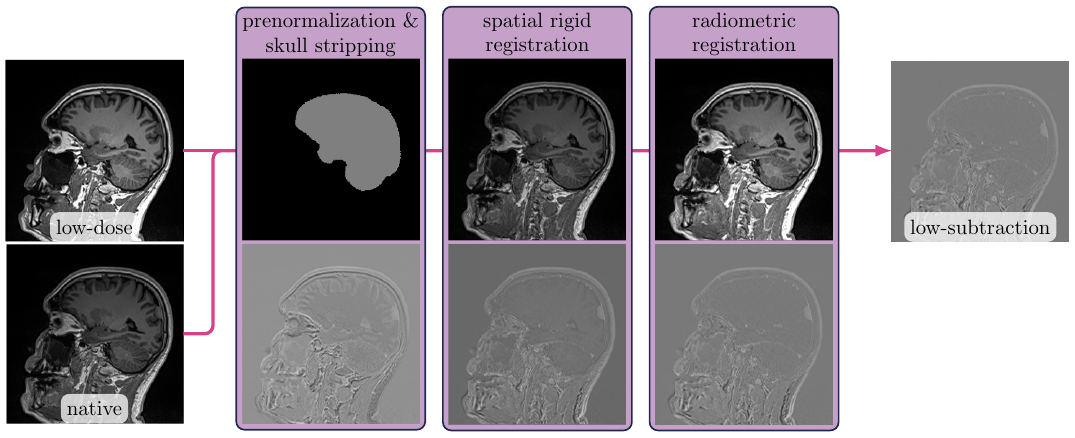}
    \caption{Illustration of the preprocessing steps for extracting the initial low-dose subtraction image~\(\tilde{\vz}_\LD\) from the pre-contrast~\(\vx_\PC\) and low-dose~\(\vx_\LD\) image. The top images in the steps show the transformed low-dose image while the bottom images visualize the effects on the subtraction image.}
    \label{fig:preprocessing}
\end{figure}

To compute the subtraction images~\(\vz_{\{\LD,\SD\}}\) from a pre-contrast~\(\vx_\PC\) and CE~\(\vx_{\{\LD,\SD\}}\) image pair, we perform the preprocessing pipeline illustrated in Figure~\ref{fig:preprocessing}.
First, the brain region~\(\vb\in\{0,1\}^n\) is extracted by an affine registration of an atlas brain towards the pre-contrast image~\(\vx_\PC\) to minimize the effects of motion artifacts from facial regions in the subsequent steps.
Second, the \texttt{itk-elastix} library is used to co-register the low-dose~\(\vx_\LD\) or standard-dose~\(\vx_\SD\) images to the corresponding pre-contrast images.
Third, each scan is coarsely normalized by mapping its $95\%$ intensity percentile of the brain region to~\(1\). 
This coarse alignment of the intensity values is refined by a radiometric registration step computing a scalar to minimize a robust distance as performed in~\cite{PiKo23}.

\begin{figure}[t]
    \centering
    \includegraphics[width=\linewidth]{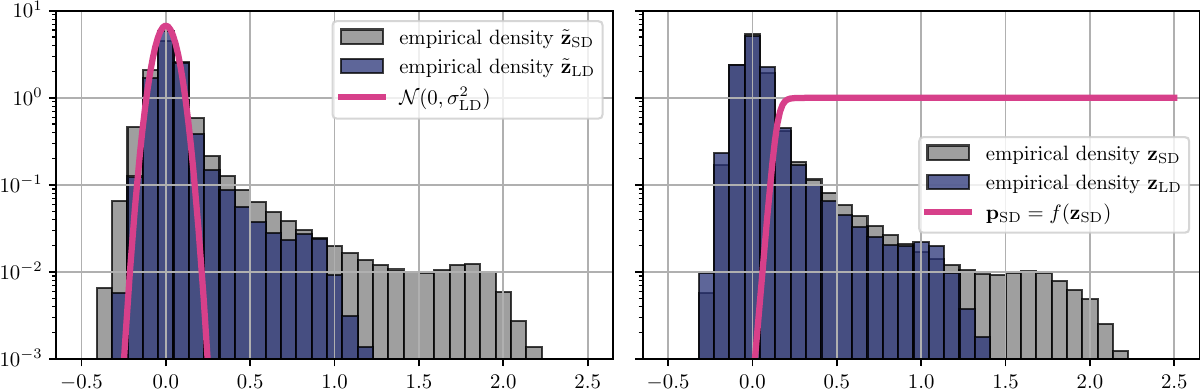}
    \caption{Visualization of the empirical densities of the different subtraction images of a prototypical case.
    The plot on the left depicts the densities of the \emph{initial} subtraction images~\(\tilde{\vz}_\LD\) and \(\tilde{\vz}_\SD\) along with the fitted Gaussian model plotted in pink.
    The associated densities of the \emph{normalized} subtraction images~\(\vz_\LD,\vz_\SD\) are shown on the right.
    The normalization ensures that both densities overlap more nicely on the right for small intensities.
    For larger intensities, the contrast-enhancing pixels of the standard-dose subtraction images dominate.
    The pink line on the right illustrates the decision of the discriminative CE signal extraction function~\eqref{eq:CESignalExtraction}.}
    \label{fig:noiseEstimation}
\end{figure}

Let us denote these initial subtraction images by~\(\tilde{\vz}_{\{\LD,\SD\}}\).
We further processed them to normalize the noise level within the images, which is often spatially varying in parallel imaging~\cite{AjVe16}.
Since the contrast signal is positive in the subtraction images, we use the negative values in the brain region to roughly estimate the noise levels~\(\sigma_\LD,\sigma_\SD\) of the low-dose and standard-dose subtraction image, respectively.
Then, we mask out potential contrast-enhancing pixels and artifacts by removing pixels that are outside of~\(\pm2\sigma_{\{\mathrm{LD},\mathrm{SD}\}}\).
On this masked subtraction image, we estimate the local mean~\(\vm_{\{\LD,\SD\}}\in\R^n\) and standard deviation~\(\vs_{\{\LD,\SD\}}\in\R^n\) using large Gaussian filter kernels with a standard deviation of \(16\)~pixels.
These local noise characteristics are then used to correct field-inhomogeneities and normalize the noise level within the subtraction images, such that
\begin{align}\label{eq:ContrastNormalization}
\vz_\LD \coloneqq \frac{\tilde{\vz}_\LD - \vm_\LD}{\vs_\LD}\sigma_\LD \approx \sigma_\LD\frac{\tilde{\vz}_\SD - \vm_\SD}{\vs_\SD} \eqqcolon \vz_\SD.
\end{align}
Note that we multiply both sides with the estimated noise level of the low-dose subtraction image~\(\sigma_\LD\) to homogenize the noise characteristics in both subtraction images.
Therefore, the two subtraction images differ essentially only in the regions where GBCA is accumulated or artifacts occurred.
This effect is illustrated by comparing the different empirical densities in Figure~\ref{fig:noiseEstimation}.

\subsection{Contrast signal extraction}

The results of the previous preprocessing steps are subtraction images~\(\vz_\LD\) and \(\vz_\SD\) after administering a low or standard dose, respectively.
These images have approximately the same noise characteristics due to normalization.

Next, we extract the \emph{contrast signal} image~\(\vy_\SD\) from the standard-dose subtraction image~\(\tilde{\vz}_\SD\).
This image serves as the ground truth for training the CNN.
To prevent the CNN from learning to synthesize noise, we aim for noise-free target images.
To this end, we use a simple discriminative model to compute a CE mask~\
\begin{align} \label{eq:CESignalExtraction}
    \vp_\SD=f(\vz_\SD)\coloneqq\sigmoid(w \vz_\SD + b) \in[0,1]^n,
\end{align}
where the parameters~\(w,b\in\R\) of the pixel-wise function~$f$ are chosen such that \(f(\sigma_\LD)=0.01\) and \(f(4\sigma_\LD)=0.99\).
These parameters represent a trade-off between noise suppression and preservation of faint CE signals.
The pink line on the right of Figure~\ref{fig:noiseEstimation} visualizes the resulting discriminative model.
This model quickly saturates to 1 and thereby preserves the CE signal for a wide range.
Finally, the target images are computed by
\[
    \vy_\SD = \vp_\SD\odot\vz_\SD,
\]
where \(\odot\) denotes the Hadamard product.

\begin{figure}[tb]
    \centering
    \includegraphics[width=\linewidth]{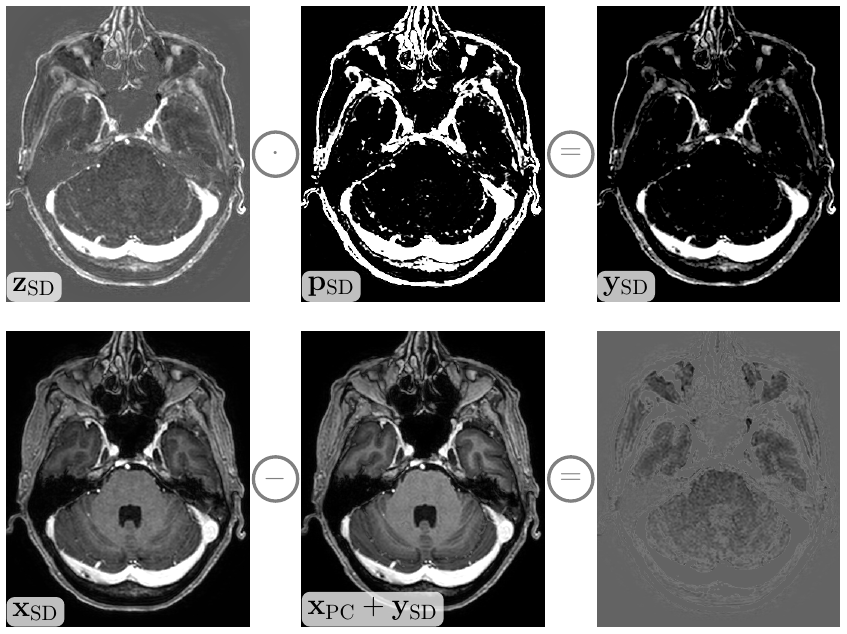}
    \caption{The top row depicts the contrast signal extraction from the standard-dose subtraction image~\(\vz_\SD\). The pixel-wise multiplication with the CE mask~\(\vp_\SD\) yields the target contrast signal image~\(\vy_\SD\).
    The bottom row visualizes the differences between the standard-dose image~\(\vx_\SD\) and the standard-dose image implicitly generated by adding the contrast signal to the pre-contrast image~\(\vx_\PC+\vy_\SD\).
    Note that both standard-dose images only differ by the noise in non-enhancing regions.}
    \label{fig:TargetGeneration}
\end{figure}

Figure~\ref{fig:TargetGeneration} illustrates the contrast signal extraction steps for a prototypical case.
As can be seen by the contrast signal image~\(\vy_\SD\) on the right of the first row, the noise is almost completely suppressed, while the CE signal is preserved.
The second row compares the standard-dose image~\(\vx_\SD\), utilized as ground truth in previous approaches, and our implicit standard-dose image generated by adding the CE signal to the pre-contrast image, i.e. \(\vx_\PC+\vy_\SD\).
These images essentially differ only in the non-enhancing regions due to noise and inhomogeneities of the bias field.
By using the contrast signal~\(\vy_\SD\) as ground truth, we avoid learning to predict the degradations shown in the bottom-right image. 

\subsection{CNN architecture}

The target of the CNN is to suppress noise and artifacts within the low-dose subtraction image~\(\vz_\LD\) and enhance the contrast signal to predict a synthetic contrast signal image~\(\hat{\vy}_\SD\) resembling the target contrast signal~\(\vy_\SD\).
Besides \(\vz_\LD\), we feed the low-dose image~\(\vx_\LD\) and the local standard deviation~\(\vs_\LD\) into the model for better distinction of contrast signal from noise and artifacts.

As sketched in Figure~\ref{fig:overview}, we use a U-Net-type architecture with residual blocks, which has proven effective in various image processing tasks~\cite{ZhLi21DRUnet}.
We adopt only 3D local operators (convolution, pooling, interpolation, etc.) in its design since contrast enhancement is a local phenomenon and the acquisition orientation is arbitrary in clinical practice.
In addition, we perform an isotropic regridding if required.
For downsampling, we use strided convolutions and tri-linear interpolation for upsampling.
At each scale, we apply two conditional residual blocks (CRBs) before down- and after upsampling.

To fuse the upsampled features with the skip connections, we propose a gated upsampling block.
Let \(\vf_u\in\R^m\) be the trilinearly upsampled features, \(\vf_s\in\R^m\) the features from the skip connection, and \(\vf_{u/s}=(\vf_u\ \vf_s)^\top\in\R^{2m}\).
Then, the fusion block combines the features by the gating mechanism
\[
\vf = \sigmoid(\widetilde{\mW}_1 \vf_{u/s})\odot \mW_1\vf_u + \sigmoid(\widetilde{\mW}_2 \vf_{u/s})\odot \mW_2\vf_s,
\]
where \(\widetilde{\mW}_{\{1,2\}}\in\R^{2m\times m}\) and \(\mW_{\{1,2\}}\in\R^{m\times m}\) are learned \(1\times1\) convolutions.
Thereby, the block can adaptively select or combine both features.

As discussed before, the overall contrast enhancement strength depends on various physical factors such as the field strength~\(B\), the contrast agent relaxivity~\(r_B\), or the relative administered dose~\(d_\LD\).
To condition the CNN on these metadata, we stack these scalars along with the low-dose noise level estimate~\(\sigma_\LD\) into a condition vector~\(\vc\in\R^4\).
This vector is processed by a two-layer fully connected network and the resulting embedding~\(\ve\in\R^{128}\) is injected into each CRB using projection matrices.
For the CRB design, we follow the approach of Ho et al.~\cite{HoJa20}, which has proven to be effective for diffusion-based generative models.
In detail, the conditional information is added as a bias term along the channel dimension in each CRB.
Thereby, the model selects the information that is relevant during training.
The exact model configuration can be found in the supplementary material or the associated github repository (released upon acceptance).

\subsection{Training and inference}
\label{sec:TrainingInference}

Given a dataset~\(\mathtt{DS} = (\vx_\PC^i,\vx_\LD^i,\vx_\SD^i,d_\LD^i, B^i,r_B^i)_{i=1}^N\), 
we learn the parameters~\(\theta\) of the CNN for contrast signal extraction and enhancement by minimizing the loss
\[
\mathcal{L}(\theta) \coloneqq \sum_{i=1}^{N} \ell(\hat{\vy}_\SD^i(\theta), \vy_\SD^i),
\]
where the sample-wise loss reads as
\[
\ell(\hat{\vy}_\SD, \vy_\SD) = \sum_{j=1}^n \left([\vb]_j + \lambda_1 + \lambda_2 [\vp_\SD]_j\right) d\left([\hat{\vy}_\SD]_j, [\vy_\SD]_j\right).
\]
Here $\vb\in\{0,1\}^n$ is the registered brain mask of the atlas, \(\vp_\SD\in[0,1]^n\) is the CE mask, and \(d: \R\times\R\to\R_+\) is the Huber loss function with \(\epsilon=0.1\).
The hyperparameters \(\lambda_1,\lambda_2>0\) control the weight on regions outside the brain mask and contrast-enhancing areas, respectively.
Note that we did not explicitly model any artifacts that typically occur in MRI due to, for instance, patient motion.
These artifacts are typically scarce in the dataset and not correlated in the different images.
Thus, the loss function will not promote any CE in these regions.

To better match the standard-dose images~\(\vx_\SD\), which are not noise normalized, we revert the noise normalization~\eqref{eq:ContrastNormalization} at inference.
Let~\(\hat{\vy}_\SD\) denote the CNN's noise-normalized output.
Then, the corresponding reverted output is defined as
\[
\tilde{\vy}_\SD = \gamma \vs_\LD \odot \hat{\vy}_\SD,
\]
where~\(\gamma>0\) accounts for different scaling factors in the brain on the training set and is computed by
\[
\gamma = \frac{1}{N}\sum_{i=1}^N \frac{\langle\vb^i,\vs_\SD^i\rangle}{\langle\vb^i,\vs_\LD^i\rangle}.
\]
Then, the synthesized standard-dose and enhanced images are given by \(\tilde{\vx}_\SD = \vx_\PC + \tilde{\vy}_\SD,\ \tilde{\vx}_{\SD+} = \vx_\LD + \tilde{\vy}_\SD\), respectively.



\section{Numerical results}

In this section, we evaluate our proposed approach quantitatively and qualitatively for synthetic and real data from multiple clinical sites.
In particular, we focus on lesions and metastases, as previous approaches have struggled to enhance these small and often subtle pathologies properly.

For the evaluation, we considered the recent approaches of Pasumarthi et al.~\cite{PaJo21} and Ammari et al.~\cite{AmBo22}.
We denote their methods as Pa-2.5D and Am-3D, respectively.
We used the available source code of Ammari et al., while we reimplemented the method of Pasumarthi et al.
To train both models, we used the hyperparameters reported in their respective papers.
Our models were trained for $2\cdot10^5$ iterations with a batch size of 16 and patches of size $96^3$ on two A40 GPUs, which took approximately 4 days each.
The hyperparameters for the loss function are set to $(\lambda_1, \lambda_2) = (0.01, 1)$ to focus on contrast-enhancing regions in the brain.
The resulting loss was optimized using Adam~\cite{KiBa15} with a learning rate of $10^{-4}$ and momentum parameters \((0.9,\, 0.999)\).
During training the learning rate was annealed to $10^{-6}$ using a cosine schedule.
To limit the influence of artifacts induced by the boundary handling, we used an 8-pixel margin for evaluating the loss.

To exclude errors due to intensity shifts in the quantitative evaluation, we registered the outputs of Am-3D and Pa-2.5D radiometrically to the pre-contrast images by minimizing the difference in non-CE regions in the brain as in~\cite{PiKo23}.
We consider the peak-signal-to-noise ratio (PSNR) evaluated on entire images PSNR\(_I\), inside the brain PSNR\(_B\), or individual lesions PSNR\(_L\) to assess the general image quality.
These regions are defined by the associated brain and lesion segmentation masks.
Furthermore, the CE of each lesion is assessed by computing the mean and maximal relative enhancement defined as
\[
\overline{c}(\vx) = \frac{\sum_{i\in L}[\vx-\vx_\PC]_i}{\sum_{i\in L}[\vx_\SD-\vx_\PC]_i},\,
\widehat{c}(\vx) = \frac{\max_{i\in L}[\vx-\vx_\PC]_i}{\max_{i\in L}[\vx_\SD-\vx_\PC]_i}.
\]
Here \(L\), is the subset of pixels included in the segmentation mask of the particular lesion.

\subsection{Datasets}

\subsubsection{Synthetic low-dose METS}

As a first dataset, we generated a synthetic low-dose brain metastasis (SLD-METS) dataset to ease data access and reproducibility.
To this end, we used the openly available code of Pinetz et al.~\cite{PiKo23} to synthesize low-dose images~\(\vx_\LD\) from pre-contrast and standard-dose image pairs~\((\vx_\PC,\vx_\SD)\) of the BraTS-METS dataset~\cite{MoJa23}.
We selected this dataset since it includes various small lesions, which are challenging for dose reduction techniques~\cite{HaPi23}.
To match the dose to the dominant one of the subsequent real low-dose dataset, we generated 
low-dose scans at a dose of \(33\%\) of the standard dose for all \(238\) samples in the BraTS-METS training set.
We randomly divided the resulting set further such that \(190\) samples were used for training and the remaining \(48\) for testing.
Note that the data of the BraTS-METS dataset does not include any meta information such as the field strength of the scanner or the administered contrast agent type.

\subsubsection{Real low-dose}

At the University Hospital Bonn, we prospectively collected \(761\) scans with multiple contrast agent doses using Ingenia \(1.5/3\)T and Achieva \(1.5/3\)T scanners (Philips Healthcare, Best, Netherlands).
For each sample, pre-contrast~\(\vx_\PC\), low-dose~\(\vx_\LD\), and standard-dose images~\(\vx_\SD\) at the same resolution were acquired in addition to the routine clinical protocol.
The GBCA Gadobutrol (\(r_{1.5}=4.6\) and \(r_3=4.5\)) or Gadoterate (\(r_{1.5}=3.9\) and \(r_3=3.4\)) were administered~\cite{ShGo15}.
For training, validation and testing, we split the dataset randomly into \(80\%/10\%/10\%\) subsets.
We denote this dataset as the real low-dose (RLD) dataset.

\begin{figure*}[t]
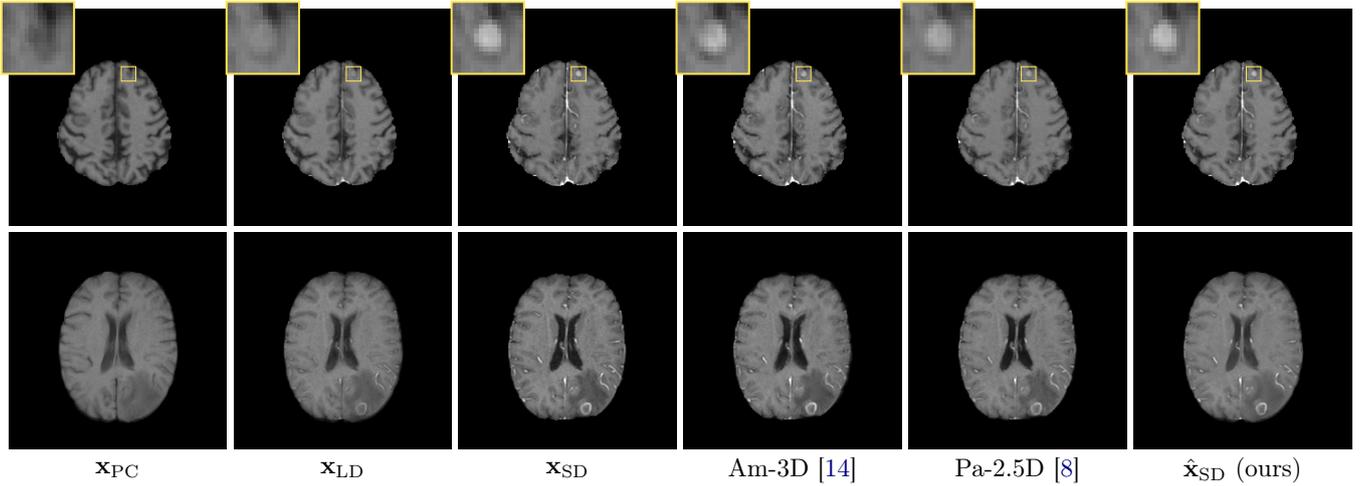

\resizebox{\linewidth}{!}{
\begin{tikzpicture}[spy using outlines={highlight,magnification=5, size=1cm}]
\tikzmath{
    \w=3;
    \dw=\w+0.1; \dh=3.1;
} 
\foreach \c/\d [count=\x from 0] in {t1n/$\vx_\PC$,t1cel/$\vx_\LD$,t1ce/$\vx_\SD$,ammari/Am-3D~\cite{AmBo22},pasu/Pa-2.5D~\cite{PaJo21},ours/$\hat{\vx}_\SD$ (ours)} {
    \node[image,rotate=180] at (\dw*\x,\dh) {\includegraphics[width=\w cm]{BraTS-MET-00104-000/\c.png}};

    \node[image] at (\dw*\x,2*\dh) {\includegraphics[width=\w cm]{BraTS-MET-00804-000/\c.png}};
    \node at (\dw*\x,\dh/2-.25) {\d};
}
\spy on (3.1*0+.15,3.1*2+.6) in node at (3.1*0-1.1,3.1*2+1.1);
\spy on (3.1*1+.15,3.1*2+.6) in node at (3.1*1-1.1,3.1*2+1.1);
\spy on (3.1*2+.15,3.1*2+.6) in node at (3.1*2-1.1,3.1*2+1.1);
\spy on (3.1*3+.15,3.1*2+.6) in node at (3.1*3-1.1,3.1*2+1.1);
\spy on (3.1*4+.15,3.1*2+.6) in node at (3.1*4-1.1,3.1*2+1.1);
\spy on (3.1*5+.15,3.1*2+.6) in node at (3.1*5-1.1,3.1*2+1.1);

\end{tikzpicture}
}
\caption{Qualitative comparison for test samples of the SLD-METS dataset.
The zooms highlight the metastasis location.
By design, our approach adds the CE signal to the input image, thereby preserving its image quality.
The input images at the top have the same resolution, while they strongly differ at the bottom. 
Thus, the image quality of our output~\(\hat{\vx}_\SD\) is poor in the bottom row and decent in the top row.}
\label{fig:bratscases}
\end{figure*}

\subsubsection{Real low-dose METS}

To investigate the generalization to different settings, we additionally collected a real low-dose metastasis (RLD-METS) dataset prospectively at an external site.
In detail, we used \(20\) samples with \(48\) metastases in total.
Here the GBCA Gadoterate was administered and the patients were scanned using $3$T Skyra or $1.5$T Aera (Siemens Healthcare, Erlangen Germany) MRI devices.
In addition, a radiology resident (three years of experience in neuroimaging) segmented all metastasis reported in the diagnostic findings.
Further details of both real-world datasets can be found in the supplementary material.

\subsection{Evaluation using synthetic low-dose images}
\label{sec:bratsEvaluation}

For this experiment, we consider the SLD-METS dataset, where each sample consists of a pre-contrast, low-dose, and standard-dose T1w image along with pre-contrast T2w and FLAIR images.
As this dataset neither includes ADC images nor scan-specific metadata, we had to adapt the methods slightly.
We replaced the additional ADC input of Am-3D with the available T2w image.
For our method, we shrank the condition to just include the estimated noise level~\(\sigma_\LD\) due to lacking information about the scanner and contrast agent.
Note that all low-dose images are synthesized for the dose~\(d_\LD=0.33\).
Thus, this dataset adheres to the assumptions of the competing DL-based dose reduction methods.
Furthermore, we discarded the noise normalization step~\eqref{eq:ContrastNormalization} of our model due to strong resolution mismatches of pre-contrast and CE images in numerous samples of the BraTS-METS dataset.

Table~\ref{tab:quant_brats} contains a quantitative evaluation of the considered dose reduction approaches.
The PSNR\(_I\) column reports the average score of the \(48\) test samples and the PSNR\(_L\) represents the mean of all \(351\) lesions in the test set.
While our approach is inferior in terms of PSNR score on the entire image, we significantly outperform all competitors on lesions.
A \(t\)-test of our approach and the closest competitor yielded the \(p\)-values presented in the last row.
The worse PSNR scores on the entire images of our model originate from the low resolution of many pre-contrast images in the SLD-METS dataset compared to the available post-contrast images.
By design, our model only focuses on the accurate prediction of the CE signal.
Thus, it cannot change the overall image quality of the input images.
Taking this into account, the PSNR\(_I\) score of our implicit ground truth~\(\vx_\PC+\vy_\SD\) is \(42.48\)dB, which is the upper bound for our approach and below the \(44.02\)dB of Ps-2.5D.
This can also be seen in the qualitative comparison depicted in the bottom row of Figure~\ref{fig:bratscases}.
The image quality of \(\vx_\PC\) is worse compared to \(\vx_\SD\).
In particular, the boundaries between the white and gray matter or the ventricles are hardly visible in \(\vx_\PC\) as it was acquired with a lower in-plane resolution.
In contrast to Am-3D and Pa-2.5D, our approach cannot overcome this lack of resolution.
Nevertheless, the CE signal is also well captured by our approach.

\begin{table}
\caption{Quantitative comparison on the SLD-METS dataset.}
\label{tab:quant_brats}
\centering
\resizebox{\linewidth}{!}{
\begin{tabular}{l | c | c | c | c}
\toprule
Model &    PSNR\(_I\) & PSNR\(_L\) & \(\overline{c}\) & \(\widehat{c}\) \\\midrule
\(\vx_\LD\) & $37.63 \pm 2.82$                & $20.52 \pm 2.90$   & $0.43\pm 0.04$ &          $0.45 \pm 0.04$           \\
Am-3D~\cite{AmBo22} & $42.74 \pm 2.62$        & $27.41 \pm 4.74$             & $0.84 \pm 0.14$ & $0.90 \pm 0.08$           \\
Pa-2.5D~\cite{PaJo21} & $\mathbf{44.02}\pm 1.94$ & $29.96 \pm 3.74$         & $0.87 \pm 0.10$ &    $0.88 \pm 0.08$     \\
\multirow[t]{2}{*}{\(\hat{\vx}_\SD\) (ours)} & $40.95\pm 4.88$  & $\mathbf{32.89} \pm 4.34$     & $\mathbf{0.98} \pm 0.05$ & $\mathbf{0.97} \pm 0.04$  \\ 
    & $p < 0.001$ & $p < 0.001$ & $p < 0.001$ & $p < 0.001$ \\
\bottomrule
\end{tabular}
}
\end{table}

The average mean and maximal relative enhancement across the \(351\) lesions is listed in the \(\overline{c}\) and \(\widehat{c}\) columns of Table~\ref{tab:quant_brats}.
Our approach significantly improves the predicted CE strength compared to Am-3D and Pa-2.5D.
Note that \(\overline{c}\) is not only sensitive to the CE strength (as \(\widehat{c}\)), but also to a lesion's internal morphology and border delineation.
A qualitative comparison of a typical lesion is visualized in the zooms at the top row of Figure~\ref{fig:bratscases}.
While Am-3D struggles to predict the internal morphology of the lesions, the CE strength of Pa-2.5D is too low.
In contrast, our approach predicts the right contrast strength, morphology, and boundary of the lesion.

\subsection{Evaluation on real low-dose images}

\begin{figure*}[t]
\resizebox{\linewidth}{!}{
\begin{tikzpicture}[spy using outlines={highlight,magnification=5, size=1cm}]
\tikzmath{
    \w=3;
    \dw=\w+0.1; \dh=3.1;
} 
\foreach[evaluate={\ptx=\x==3?-1.15:-.68;\pty=\x==3?-.6:+.3}] \c/\d [count=\x from 0] in {t1n/$\vx_\PC$,t1cel/$\vx_\LD$,t1ce/$\vx_\SD$,t1_ammari/Am-3D~\cite{AmBo22},t1_pasu/Pa-2.5D~\cite{PaJo21},t1_ours/$\tilde{\vx}_\SD$ (ours)}{
    \foreach \e [count=\y from 0] in {VC_000000118,VC_000000691}
        \node[image,anchor=north] at (\dw*\x,\dh*\y) {\includegraphics[width=\w cm]{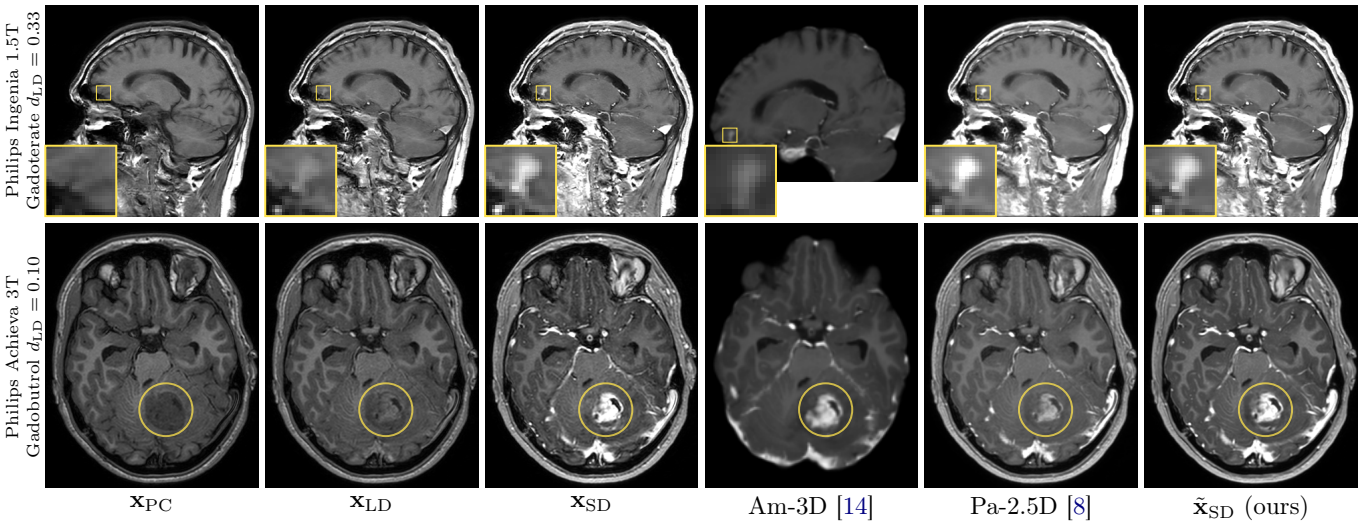}};
    \node[anchor=north] at (\dw*\x,-3.75) {\d};
    \node[marker, minimum size=.75cm] at (\dw*\x+.2,0-2.65) {};
}

\node[rotate=90,font=\scriptsize,align=center,inner sep=0] at (-1.8,-1.9) {Philips Achieva 3T\\ Gadobutrol \(d_\LD=0.10\)};
\node[rotate=90,font=\scriptsize,align=center,inner sep=0] at (-1.8,1.6) {Philips Ingenia 1.5T\\ Gadoterate \(d_\LD=0.33\)};


\spy on (3.1*0-.68,3.5*1-1.65) in node at (3.1*0-1.,3.5*1-2.9);
\spy on (3.1*1-.68,3.5*1-1.65) in node at (3.1*1-1.,3.5*1-2.9);
\spy on (3.1*2-.68,3.5*1-1.65) in node at (3.1*2-1.,3.5*1-2.9);
\spy on (3.1*3-1.15,3.5*1-2.25) in node at (3.1*3-1.,3.5*1-2.9);
\spy on (3.1*4-.68,3.5*1-1.65) in node at (3.1*4-1.,3.5*1-2.9);
\spy on (3.1*5-.68,3.5*1-1.65) in node at (3.1*5-1.,3.5*1-2.9);

\end{tikzpicture}
}
\caption{Qualitative evaluation on the RLS dataset.
The top row depicts a \(33\%\) low-dose sample, where the yellow zoom highlights the lesion.
The bottom row shows a case with a \(10\%\) real low-dose image and the lesion is marked by the yellow circles.}
\label{fig:realcases}
\end{figure*}

Next, Am-3D, Pa-2.5D, and our approach trained on the RLD dataset are evaluated on the \(77\) RLD test images.
Here, ADC images are additionally fed into Am-3D as required~\cite{AmBo22}.

The quantitative results for this dataset are depicted in Table~\ref{tab:quant_real}.
In addition to the structured similarity index (SSIM), we computed the mean absolute error in CE brain regions MAE\(_\CE\) as suggested in~\cite{PiKo23} to compare the contrast enhancement due to a lack of lesion segmentation masks for this dataset.
For Am-3D, only a PSNR score in the brain region w.r.t. their corresponding ground truth is reported as their preprocessing requires skull stripping and a non-invertible contrast normalization step.
The results in Table~\ref{tab:quant_real} show that our approach significantly outperforms previous approaches in terms of PSNR score in the brain and the MAE in the contrast-enhancing brain regions.
The corresponding \(p\)-values of a \(t\)-test are listed in the last row of Table~\ref{tab:quant_real}.
As a result, our approach predicts a contrast enhancement signal closer to the ground truth.
Moreover, the PSNR values for the whole image and the SSIM are also improved, although, unlike Pa-2.5D, the SSIM is not part of our loss function.
Note that this is in contrast to the PSNR\(_I\) scores for the SLD-METS dataset; however, in the RLD dataset, all T1w images are acquired with the same resolution and have the same image quality.

\begin{table}
\caption{Quantitative evaluation on the RLD dataset.}
\label{tab:quant_real}
\centering
\resizebox{\linewidth}{!}{
\begin{tabular}{l | c | c | c | c}
\toprule
Model &    PSNR\(_I\) & PSNR\(_B\) & SSIM & MAE\(_\CE\) \\\midrule
    $\vx_\LD$     & $29.32 \pm 2.27$  & $29.96 \pm 3.54$      & $.899 \pm .038$ & $0.28 \pm 0.014$ \\
Am-3D~\cite{AmBo22}     & $-$               & $30.77^\ast \pm 2.40$ & $-$ & $-$              \\
Pa-2.5D~\cite{PaJo21} & $32.45 \pm 3.33$  & $33.66 \pm 4.36$      & $.916\pm .049$   & $0.17 \pm 0.07$  \\
\multirow[t]{2}{*}{$\tilde{\vx}_\SD$ (ours)} & $\mathbf{32.63} \pm 2.64$    & $\mathbf{35.05} \pm 4.19 $  & $\mathbf{.919} \pm .038 $ & $\mathbf{0.15 \pm 0.05} $   \\
                                &    $p=0.33$                  &     $p < 0.001$ & $p=0.43$ &               $p = 0.011$  \\ \bottomrule
\multicolumn{5}{l}{\(^\ast\)Am-3D is evaluated w.r.t. its associated ground truth.}
\end{tabular}
}
\end{table}


A qualitative comparison of RLD test samples is shown in Figure~\ref{fig:realcases}.
The required skull stripping and non-linear intensity normalization of Am-3D are readily apparent.
Thus, a quantitative comparison to the target~\(\vx_\SD\) is not reasonable.
Nevertheless, the CE signal strength in pathological regions (highlighted by the yellow circles) is well visible despite too smooth output images.
The internal morphology and the border delineation also exhibit blurring.
Pa-2.5D yields better image quality but the contrast strength in pathological regions is not well captured.
In particular, the contrast signal overshoots for the \(33\%\) low-dose scan in the first row and undershoots for the \(10\%\) scan at the bottom, although the training dataset contains low-dose images at both dose levels.
Our approach yields the highest image quality and predicts the contrast enhancement more accurately due to the focus on the enhancement signal and the conditional embeddings.
As a result, our approach generates CE images preserving the image quality of the inputs by avoiding the synthesis of anatomical or noise patterns.

\subsection{Generalization on real low-dose metastases}

To analyze the generalization capabilities of the different approaches, we use the RLD-METS dataset.
In addition, we compare the performance of the models trained on the SLD-METS and RLD training sets.
To apply the models trained on the SLD-METS dataset, we resampled all images to \(1\text{mm}^3\) and preprocessed them using the \texttt{CaPTK}\footnote{\url{https://cbica.github.io/CaPTk/}} toolbox according to BraTS requirements.
Note that this preprocessing failed on \(4\) test samples with \(5\) lesions in total, which we excluded for the evaluation of the models trained on SLD-METS.
Further, we exclude Am-3D~\cite{AmBo22} trained on RLD due to limited comparability caused by the differences in preprocessing.
Consequently, the images were not skull-stripped and processed on the acquired resolution for the Pa-2.5D and our model trained on the RLD dataset.

\begin{table}
\caption{Quantitative metrics on the RLD-METS dataset. The models in the top row were trained on the SLD-METS dataset and the bottom rows were trained on the RLS dataset.}
\label{tab:quant_synt_ours}
\centering
\resizebox{\linewidth}{!}{
\begin{tabular}{ll | c | c | c | c }
\toprule
&Model        & PSNR\(_B\)       & PSNR\(_L\)        & $\overline{c}$ & $\widehat{c}$  \\\midrule
\multirow{5}{*}{\rotatebox[origin=c]{90}{{\scriptsize SLD-METS}}}&$\vx_\LD$           &      $27.37 \pm 1.07$ & $19.70 \pm 3.01$  & $0.36 \pm 0.32$ & $0.46 \pm 0.10$             \\
    &Am-3D~\cite{AmBo22}          &  $26.50 \pm 1.81$             & $23.41 \pm 2.53$  & $0.82 \pm 0.21$           & $\mathbf{0.85} \pm 0.19$             \\
    &Pa-2.5D~\cite{PaJo21}       & $28.27 \pm 2.10$ & $25.75 \pm 2.44$  & $0.77 \pm 0.22$           & $0.82 \pm 0.18$             \\
    &\multirow[t]{2}{*}{$\tilde{\vx}_\SD$ (ours)} & $\mathbf{29.21} \pm 2.26$ & $\mathbf{26.28} \pm 2.78$ & $\mathbf{0.84}\pm 0.16$&$0.84\pm 0.17$ \\ 
    && $p < 0.001$ & $p = 0.186$ & $p = 0.64$ & $p=0.49$ \\
    \midrule
\multirow{5}{*}{\rotatebox[origin=c]{90}{{\scriptsize RLD}}}&$\vx_\LD$ & $31.13 \pm 1.76$ & $23.52 \pm 4.32$  & $0.39 \pm 0.09$           & $0.54 \pm 0.12$             \\
    &Pa-2.5D~\cite{PaJo21}  & $32.82 \pm 1.98$ & $27.39 \pm 3.47$  & $0.58\pm 0.38$           & $0.70 \pm 0.27$             \\
    &\multirow[t]{2}{*}{$\tilde{\vx}_\SD$ (ours)}& $\mathbf{33.13} \pm 2.45$ & $\mathbf{28.38} \pm 3.98$ & $0.78 \pm 0.34$ & $0.83 \pm  0.36$    \\
    &                        & $p=0.187$      &  $p=0.047$      & $p < 0.001$  &   $p=0.004$                   \\ 
    &$\tilde{\vx}_{\SD+}$ (ours)& $-$              & $-$               & $\mathbf{1.17} \pm 0.42$           &  $\mathbf{1.33} \pm 0.42$            \\ \bottomrule 
\end{tabular}
}
\end{table}

\begin{figure*}
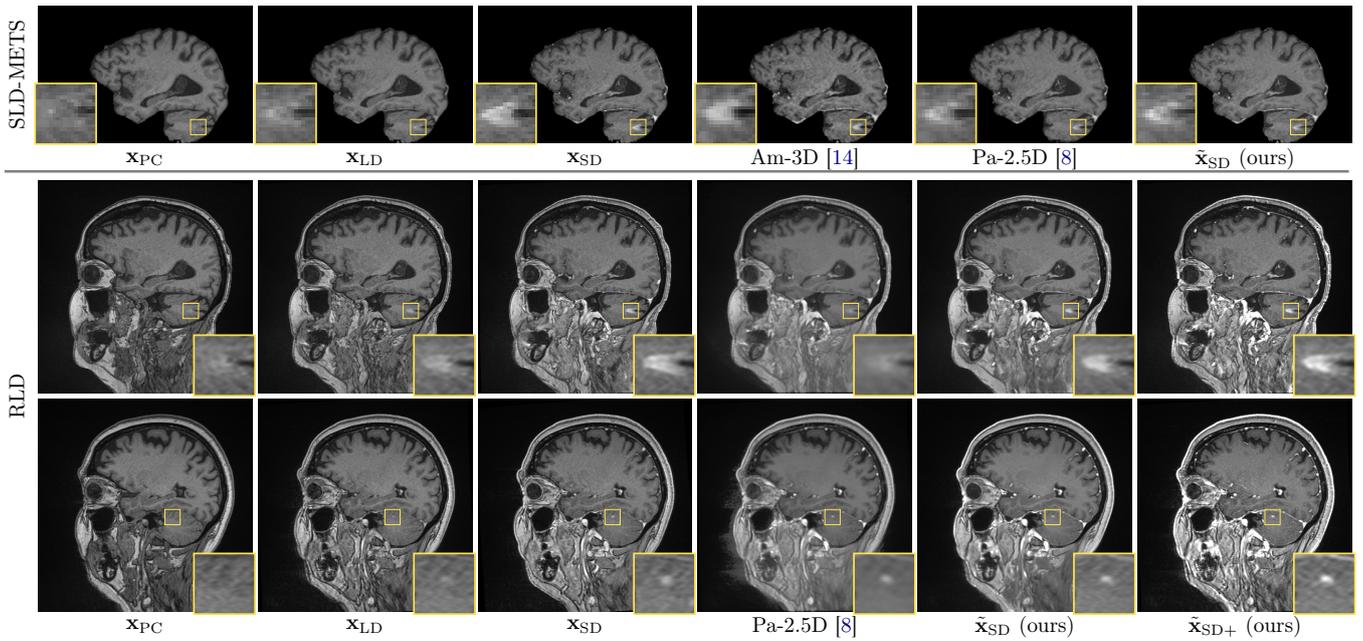

\resizebox{\linewidth}{!}{
\begin{tikzpicture}[spy using outlines={highlight,magnification=4, size=1cm}]
\tikzmath{
    \w=3.5;
    \dw=\w+0.1;
} 

\node[rotate=90] at (-2.1,3.5) {SLD-METS};
\draw[gray, very thick] (-2.3,1.9) -- +(22,0);

\foreach \c/\d [count=\x from 0] in {t1n/$\vx_\PC$,t1cel/$\vx_\LD$,t1ce/$\vx_\SD$,ammari/Am-3D~\cite{AmBo22},pasu/Pa-2.5D~\cite{PaJo21},ours/$\tilde{\vx}_\SD$ (ours)} {
    \node[inner sep=0] at (\dw*\x,3.5) {\includegraphics[width=\w cm]{SC_0146_brats/\c.png}};
    \node at (\dw*\x,2.1) {\d};
}

\spy on (3.6*0+.87,3.5-.87) in node at (3.6*0-1.3,3.5-.65);
\spy on (3.6*1+.87,3.5-.87) in node at (3.6*1-1.3,3.5-.65);
\spy on (3.6*2+.87,3.5-.87) in node at (3.6*2-1.3,3.5-.65);
\spy on (3.6*3+.87,3.5-.87) in node at (3.6*3-1.3,3.5-.65);
\spy on (3.6*4+.87,3.5-.87) in node at (3.6*4-1.3,3.5-.65);
\spy on (3.6*5+.87,3.5-.87) in node at (3.6*5-1.3,3.5-.65);

\node[rotate=90] at (-2.1,-3.6/2) {RLD};

\foreach \c/\d [count=\x from 0] in {t1n/$\vx_\PC$,t1cel/$\vx_\LD$,t1ce/$\vx_\SD$,t1_pasu/Pa-2.5D~\cite{PaJo21},t1_ours/$\tilde{\vx}_\SD$ (ours), t1_ours_plus/$\tilde{\vx}_{\SD+}$ (ours)} {
    \node[inner sep=0] at (\dw*\x,0) {\includegraphics[width=\w cm]{SC_0146/\c.png}};
}

\spy on (3.6*0+.75,0-.4) in node at (3.6*0+1.3,0-1.3);
\spy on (3.6*1+.75,0-.4) in node at (3.6*1+1.3,0-1.3);
\spy on (3.6*2+.75,0-.4) in node at (3.6*2+1.3,0-1.3);
\spy on (3.6*3+.75,0-.4) in node at (3.6*3+1.3,0-1.3);
\spy on (3.6*4+.75,0-.4) in node at (3.6*4+1.3,0-1.3);
\spy on (3.6*5+.75,0-.4) in node at (3.6*5+1.3,0-1.3);

\foreach \c/\d [count=\x from 0] in {t1n/$\vx_\PC$,t1cel/$\vx_\LD$,t1ce/$\vx_\SD$,t1_pasu/Pa-2.5D~\cite{PaJo21},t1_ours/$\tilde{\vx}_\SD$ (ours), t1_ours_plus/$\tilde{\vx}_{\SD+}$ (ours)} {
    \node[inner sep=0] at (\dw*\x,-3.6) {\includegraphics[width=\w cm]{SC_0082/\c.png}};
    \node at (\dw*\x,-5.6) {\d};
}

\spy on (3.6*0+.45,-3.6-.2) in node at (3.6*0+1.3,-3.6-1.3);
\spy on (3.6*1+.45,-3.6-.2) in node at (3.6*1+1.3,-3.6-1.3);
\spy on (3.6*2+.45,-3.6-.2) in node at (3.6*2+1.3,-3.6-1.3);
\spy on (3.6*3+.45,-3.6-.2) in node at (3.6*3+1.3,-3.6-1.3);
\spy on (3.6*4+.45,-3.6-.2) in node at (3.6*4+1.3,-3.6-1.3);
\spy on (3.6*5+.45,-3.6-.2) in node at (3.6*5+1.3,-3.6-1.3);

\end{tikzpicture}
}
\caption{Qualitative comparison on the RLD-METS dataset.
The top row depicts the outputs for the models trained on the SLD-METS dataset. The bottom two rows show the outputs of the models trained on the RLD dataset.
The samples in the first and second rows are identical.
All images were acquired by a Siemens Aera 1.5T scanner using \(d_\LD=0.33\). The lesions are highlighted by yellow zooms.}
\label{fig:metacases}
\end{figure*}

Table~\ref{tab:quant_synt_ours} lists the resulting PSNR and relative contrast enhancement scores.
The top rows show the metrics for the models trained on the SLD-METS dataset, and the bottom rows those of the models trained using RLD.
Our models yield the best PSNR scores in the brain regions and for lesions.
As before, we computed \(t\)-test of our models with its closest competitors,
The resulting \(p\)-values are shown below the corresponding \(\tilde{\vx}_\SD\) (our) rows.
For our SLD-METS model, the PSNR improvement is significant in the brain region, while our RLD model significantly outperforms the other models for PSNR\(_L\).
The outputs of Am-3D yield the best mean and maximal contrast enhancement~\(\overline{c}/\widehat{c}\) for the models trained on the SLD-METS dataset.
However, the improvement is not significant compared to our approach.
For the models trained on the RLD dataset, our approach is superior by a clear margin. 
In general, there is a consistent improvement in image quality for all models when trained on real low-dose (RLD) images.
Interestingly, the relative contrast enhancement scores decrease for all methods, which might be because the SLD-METS dataset only contains pathological images.

A qualitative comparison of the different models is visualized in Figure~\ref{fig:metacases}.
The first and second rows highlight the same lesion of a case.
The first row shows the outputs of the models trained on the SLD-METS dataset, using the \texttt{CaPTK} preprocessing of BraTS.
Am-3D\cite{AmBo22} enhances the lesion the strongest but also emphasizes streaking artifacts, resulting in degraded image quality.
In contrast, the middle row depicts the results of the models trained on the RLD dataset.
Comparing the zooms of the first and second rows, we see that the RLD models generate a better image quality due to operating on higher resolutions.
In addition, the output of Pa-2.5D~\cite{PaJo21} in the second row shows a hardly enhanced lesion, which might be due to multiple dose levels in the RLD dataset.
The last row presents the case in which our model performed the worst in terms of PSNR\(_L\) compared to Pa-2.5D.
Nevertheless, the lesion is also clearly visible in our output.
Finally, we highlight that by adding the predicted contrast signal~\(\tilde{\vy}_\LD\) to the low-dose image we obtain images~\(\tilde{\vx}_\SD\) with a stronger CE.
The associated mean and maximal relative enhancements listed at the bottom of Table~\ref{tab:quant_synt_ours} confirm that \(\tilde{\vx}_{\SD+}\) exceeds the CE of standard-dose images. 


\subsection{Embedding analysis}

\begin{figure}[t]
\centering
\includegraphics[width=\linewidth]{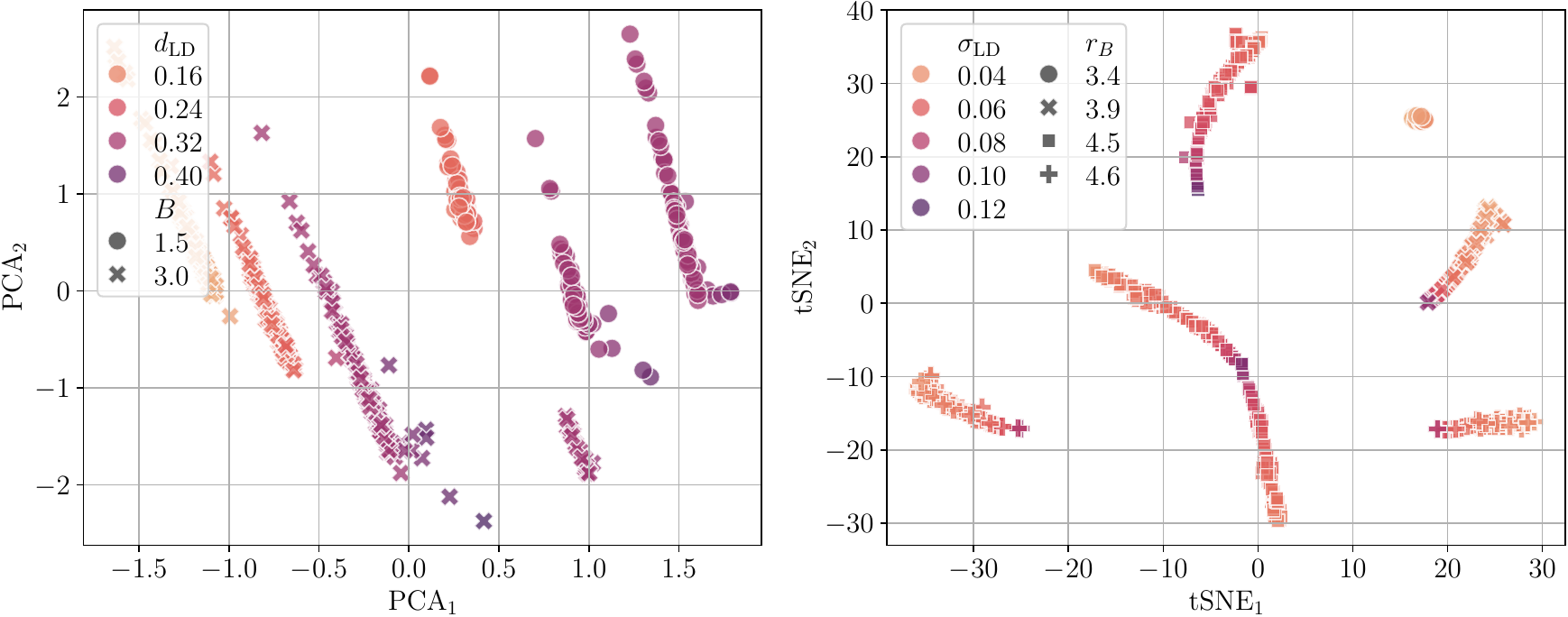}
\caption{Visualization of the embedding vectors on the RLD training set.
The PCA plot on the left shows the dependency on the field strength~\(B\) and dose~\(d_\LD\), while the tSNE plot on the right resolves the effect of the contrast agent relaxivity~\(r_B\) and the noise level~\(\sigma_\LD\) on the embedding.}
\label{fig:EmbeddingAnaylsis}
\end{figure}

Next, we perform a qualitative analysis of the embedding.
To this end, we extracted the embedding vectors~\((\ve^i)_{i=1}^N\) for all training samples of the RLD-dataset.
Using the principal component analysis (PCA) and the t-distributed stochastic neighbor embedding (tSNE)~\cite{VaHi08}, we computed mappings from the \(128\)-dimensional embedding space into 2D, shown in Figure~\ref{fig:EmbeddingAnaylsis}.
The PCA plot on the left demonstrates an almost linear dependency of the embedding on the field strength~\(B\in\{1.5,3\}\) and the administered dose~\(d_\LD\in(0,1)\).
In contrast, the tSNE plot on the right shows that the embedding vectors form clusters depending on the contrast agent's relaxivity~\(r_B\in\R_+\).
Furthermore, the standard deviation~\(\sigma_\LD\) smoothly changes in each cluster.
Thus, each acquisition parameter affects the embedding, which in turn controls our models' output.

\section{Conclusion}

In this work, we presented a novel approach for GBCA dose reduction using subtraction images.
We demonstrated that our conditional CNN denoises and enhances initially degraded subtraction images of low-dose and pre-contrast images to extract the contrast signal.
Moreover, we demonstrated that adding the resulting contrast signal to the pre-contrast or low-dose image yields realistic CE images, which can be enhanced beyond the standard dose.
The extensive numerical evaluations on synthetic and real low-dose images showed that our approach significantly outperforms the state-of-the-art, particularly for lesions in the brain.
In addition, embedding meta-information of the acquired images such as the injected contrast agent type and dose, or the scanner's field strength is beneficial for predicting the enhancement strength more accurately.
In the future, we aim to evaluate our approach clinically to reduce GBCA administration and simultaneously increase the visibility of pathologies.


\printbibliography
\clearpage

\section*{Supplementary material}

\subsection*{CNN architecture details}

\begin{figure}[!h]
\centering
\includegraphics[width=\linewidth]{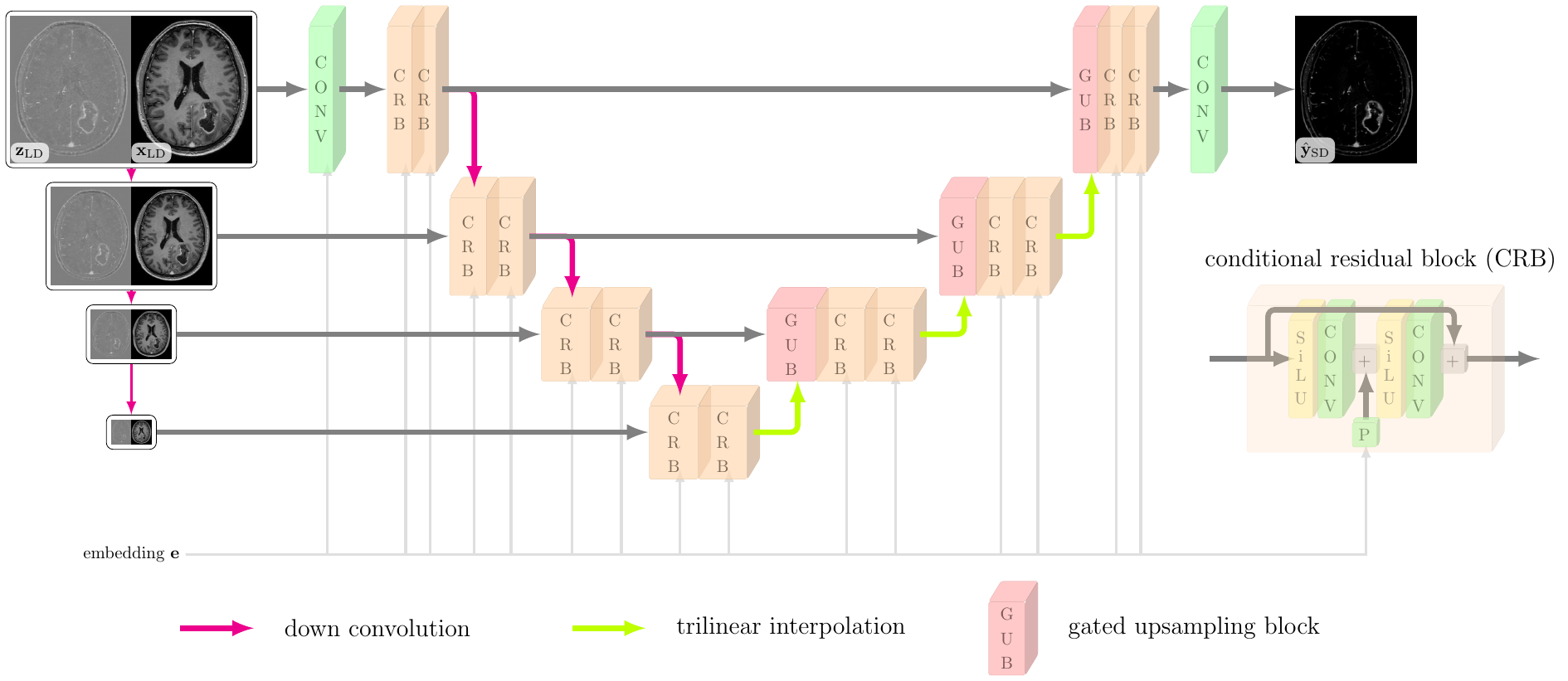}
\caption{Illustration of the conditional U-net-like CNN. The input images\((\vz_\LD,\vx_\LD)\) are downscaled and then fed into the conditional residual blocks (CRBs) of the corresponding scale. The embedding~\(\ve\) is projected to the corresponding number of channels in each CRB and added to the residual path. To fuse features of different scales, the gated upsampling blocks implement a pixel-wise gating approach.}
\label{fig:model}
\end{figure}

Here, we provide additional details of the used CNN architecture, illustrated in Figure~\ref{fig:model}.
The model operates on 4 scales and each image/feature channel is downsampled using down convolution.
More precisely, we use pre-defined \(5\times5\times5\) smoothing kernels according to Zhang~\cite{Zh19} to promote shift-invariance.
Further, we upsample lower-resolution features using trilinear interpolation operators.
We use \(3\times3\times3\) convolution filters in every convolution layer and extract on the finest scale \(32\) feature channels.
The number of features is doubled after every downsampling operation.
For each non-linearity, we use the SiLU activation function within the network.
At the model output, we apply a ReLU activation function to ensure the positivity of the predicted contrast enhancement.

The main building blocks of the architecture are conditional residual blocks (CRBs).
Each of these blocks is conditioned onto the embedding by projecting the embedding vector the corresponding number of feature channels before adding the result to the residual path.
This design is adapted from Ho et al.~\cite{HoJa20} to account for 3D features.
In addition, we removed the normalization layers.

We introduced the gated upsampling block to merge upsampled features with the skip connections.
Let \(\vf_u\in\R^m\) be the trilinearly upsampled features, \(\vf_s\in\R^m\) the features from the skip connection, and \(\vf_{u/s}=(\vf_u\ \vf_s)^\top\in\R^{2m}\).
Then, the fusion layer combines the features by the gating mechanism
\[
\vf = \sigmoid(\widetilde{\mW}_1 \vf_{u/s})\odot \mW_1\vf_u + \sigmoid(\widetilde{\mW}_2 \vf_{u/s})\odot \mW_2\vf_s,
\]
where \(\widetilde{\mW}_{\{1,2\}}\in\R^{2m\times m}\) and \(\mW_{\{1,2\}}\in\R^{m\times m}\) are learned \(1\times1\) convolutions.
Thereby, the layer can adaptively select or combine both features.

\subsection*{Datasets}

\subsubsection*{RLD dataset}

For \(761\) subjects, three T1-weighted (T1w) images with multiple GBCA dose levels (\(d_\PC=0\), \(d_\LD\in[0.1,0.33]\), \(d_\SD=1\)) were acquired at \(2\) clinical locations within our institution on \(5\) different MRI systems and \(2\) Gadolinium-based contrast agent (GBCA) types.
MRI systems included 3T Achieva and Ingenia as well as 1.5T Achieva and Ingenia (Philips Healthcare, Best, Netherlands).
Each triplet of T1w images was acquired at the same scanner within a time frame of 30min using the same acquisition sequence.
Table~\ref{tab:rldDataset} lists the distribution of the samples to the different settings.

\begin{table}[th]
\caption{Overview of the real low-dose (RLD) dataset.}
\label{tab:rldDataset}
\centering
\resizebox{\linewidth}{!}{
\begin{tabular}{l c c c c c c c}
\toprule
System                      & Field strength & GBCA                           & Sequence &\(T_E\)    &\(T_R\)     & \(\alpha\) & Count \\ \midrule
\multirow[t]{2}{*}{Achieva} & 1.5            & \multirow[t]{2}{*}{Gadobutrol} & T1-TFE   & 3.26-3.45 & 7.23-7.55  & 8  & 178 \\
                            & 3              &                                & T1-TFE   & 2.96-3.33 & 6.61-7.22  & 8  & 435 \\
\multirow[t]{2}{*}{Ingenia} & 1.5            & \multirow[t]{2}{*}{Gadoterate} & T1-FFE   & 3.59-3.60 &15.15-15.51 & 30 & 119 \\
                            & 3              &                                & T1-TFE   & 3.69-3.89 & 7.52-7.93  & 8  & 29 \\ \midrule
Total                       &                &                                &          &           &            &    & 761 \\
\bottomrule
\end{tabular}
}
\end{table}

\subsubsection*{RLD-METS dataset}
In contrast to the RLD dataset above, this dataset was acquired at an external site.
In detail, \(20\) subjects were scanned to obtain three T1w scans with different dose levels (\(d_\PC=0\), \(d_\LD=0.33\), \(d_\SD=1\)) using the GBCA Gadoterate.
Magnets included 3T Skyra and 1.5T Aera (Siemens Healthcare, Erlangen Germany).
All images were acquired using a 3D TurboFlash pulse sequence; the corresponding parameters along with the distribution of the samples are depicted in Table~\ref{tab:rldMETSDataset}
In total, there are \(47\) lesions in the RLD-METS dataset and each sample has at least one metastasis and at most 5.
Thus, these pathologies are homogeneously distributed across the samples.

\begin{table}[h]
\caption{Overview of the real RLD-METS test set.}
\label{tab:rldMETSDataset}
\centering
\resizebox{\linewidth}{!}{
\begin{tabular}{l c c c c c}
\toprule
System & Field strength & \(T_E\) & \(T_R\)       & \(\alpha\) & Count \\ \midrule
Aera   & 1.5            & 3.15    & 1920.0-1990.0 & 10         & 11 \\
Skyra  & 3              & 2.20    & 1640.0-2000.0 & 9          & 9 \\ \midrule
Total  &                &         &               &            & 20 \\
\bottomrule
\end{tabular}
}
\end{table}

\end{document}